\documentclass[12pt]{article}
\usepackage{epsfig,amsfonts,amssymb}
\begin{document}

\newcommand{\be}{\begin{equation}}
\newcommand{\ee}{\end{equation}}
\newcommand{\bea}{\begin{eqnarray}}
\newcommand{\eea}{\end{eqnarray}}
\newcommand{\beas}{\begin{eqnarray*}}
\newcommand{\eeas}{\end{eqnarray*}}
\newcommand{\dtphi}{\dot{\varphi}}
\newcommand{\erf}{\mbox{erf}}
\newcommand{\erfc}{\mbox{erfc}}

\font\cmss=cmss10
\def\half{{1 \over 2}}
\def\identity{{\rlap{\cmss 1} \hskip 1.6pt \hbox{\cmss 1}}}
\def\laplace{{\kern1pt\vbox{\hrule height 1.2pt\hbox{\vrule width 1.2pt\hskip
  3pt\vbox{\vskip 6pt}\hskip 3pt\vrule width 0.6pt}\hrule height 0.6pt}
  \kern1pt}}
\def\scriptlap{{\kern1pt\vbox{\hrule height 0.8pt\hbox{\vrule width 0.8pt
  \hskip2pt\vbox{\vskip 4pt}\hskip 2pt\vrule width 0.4pt}\hrule height 0.4pt}
  \kern1pt}}
\def\slash#1{{\rlap{$#1$} \thinspace /}}
\def\roughly#1{\raise.3ex\hbox{$#1$\kern-.75em\lower1ex\hbox{$\sim$}}}
\def\complex{{\hbox{\cmss C} \llap{\vrule height 7.0pt
  width 0.4pt depth -.4pt \hskip 0.5 pt \phantom .}}}
\def\real{{\hbox{\cmss R} \llap{\vrule height 6.9pt width 0.4pt
  depth -.1pt \hskip 0.6 pt \phantom .}}}
\def\integer{{\rlap{\cmss Z} \hskip 1.8pt \hbox{\cmss Z}}}

\font\cmsss=cmss8
\def\C{{\hbox{\cmsss C}}}
\def\bigC{{\hbox{\cmss C}}}
\def\gs{{g^2_{YM}}}
\def\gssb{{g^2_{YM} \sqrt{\beta}}}
\def\gsb{{g^2_{YM} \beta}}
\def\gfb{{g^4_{YM} \beta}}

\baselineskip 14 pt
\parskip 12 pt

\begin{titlepage}
\begin{flushright}
{\small CU-TP-1109} \\
{\small FERMILAB-PUB-04-162-A} \\
{\small RUNHETC-2003-38}
\end{flushright}

\begin{center}

\vspace{1mm}

{\Large \bf String Windings in the Early Universe}

\vspace{1mm}

Richard Easther$^{a}$, Brian R.\ Greene$^{bc}$, Mark G.\
Jackson$^{bd}$, Daniel Kabat$^{be}$

\vspace{1mm}

{\small \sl $^a$Department of Physics} \\
{\small \sl Yale University, New Haven, CT 06520} \\
{\small \tt richard.easther@yale.edu}

{\small \sl $^b$Institute for Strings, Cosmology and Astroparticle Physics} \\
{\small \sl and Department of Physics} \\
{\small \sl Columbia University, New York, NY 10027} \\
{\small \tt greene, markj, kabat@physics.columbia.edu}

{\small \sl $^c$Department of Mathematics} \\
{\small \sl Columbia University, New York, NY 10027}

{\small \sl $^d$NASA/Fermilab Theoretical Astrophysics Group} \\
{\small \sl Fermilab, Batavia, IL 60510}

{\small \sl $^e$Department of Physics and Astronomy} \\
{\small \sl Rutgers University, Piscataway, NJ 08855-0849}

\end{center}

\noindent
We study string dynamics in the early universe.  Our motivation is the
proposal of Brandenberger and Vafa, that string winding modes may play
a key role in decompactifying three spatial dimensions.  We model the
universe as a homogeneous but anisotropic 9-torus filled with a gas of
excited strings.  We adopt initial conditions which fix the
dilaton and the volume of the torus, but otherwise assume all states
are equally likely. We study the evolution of the system both
analytically and numerically to determine the late-time behavior.  We
find that, although dynamical evolution can indeed lead to three large
spatial dimensions, such an outcome is not statistically favored.

\end{titlepage}

\section{Introduction}
An enduring challenge for string/M-theory is to provide a more
complete picture of the early universe than has been found using
conventional, point-particle approaches. To this end, a growing body
of research has studied the dynamics of strings and branes in a
cosmological setting, as opposed to the more widely investigated case
of a static background. Intriguing but as yet incomplete results have
been found for higher dimensional cosmologies and cosmologies based on
braneworlds. These include mechanisms for resolving or avoiding
cosmological singularities, and for generating subtle modifications to
the primordial microwave background power spectrum.  Further progress
on these key theoretical and observational issues, however, requires a
more refined grasp of the dynamical properties of strings and branes
when subject to extremes of temperature, density, and curvature. The
current paper provides a modest step in this direction.

The formalism we develop can, in principle, be applied to a wide range
of string cosmology questions. But following our earlier works
\cite{Easther:2002qk} \cite{Easther:2003dd}, our immediate goal is to
find a dynamical mechanism within string/M-theory that generically
gives rise to a universe with precisely three large spatial
dimensions, with all other spatial dimensions unobservably small. Such
an asymmetric dynamical evolution is perhaps the most basic task of
string/M cosmology. However, a decade and a half after the first
attempt, no satisfactory picture has yet emerged.

By way of brief history, in \cite{BrandenbergerVafa} and
\cite{TseytlinVafa} the authors made use of T-duality in a (spatially)
toroidal universe to argue that strings wound around nontrivial cycles
impede the growth rate of the spatial dimensions they wrap.  The
fastest expansion will therefore be achieved by dimensions that shed
all their winding modes through string winding/anti-winding
annihilations. Because string worldsheets are two-dimensional, pairs
of strings will generically intersect in four or fewer spacetime
dimensions, leading \cite{BrandenbergerVafa} and \cite{TseytlinVafa}
to argue that at most three spatial dimensions will shed their
windings and subsequently grow with time.  Various aspects
of this proposal have been investigated and generalized. In
\cite{Sakellariadou:1995vk}, a numerical study of a gas of strings in
a toroidal universe was carried out and the naive dimension counting
argument for string annihilations used in \cite{BrandenbergerVafa}
(2+2 = 3+1) was verified in a static background. In
\cite{Easther:2002mi}, the analysis was further extended to simply
connected toroidal orbifolds, and it was argued that pseudo-winding
modes with sufficiently long lifetimes could allow the arguments of
\cite{BrandenbergerVafa} to apply in phenomenologically relevant
backgrounds. In \cite{Alexander:2000xv}, higher dimensional branes
were included in the analysis, without invalidating the
conclusions of \cite{BrandenbergerVafa}.  Other studies of string
windings and brane gases can be found in
\cite{Cleaver:1994bw}-\cite{Campos:2004yn}.

All these works, however, fail to account for the detailed
cosmological dynamics. In \cite{Easther:2002qk} we partially addressed
this deficiency by using eleven-dimensional supergravity to study
cosmological evolution in the presence of a brane gas and found
encouraging results: for a suitable configuration of brane wrappings
suggested in \cite{Alexander:2000xv} and \cite{Easther:2002qk} and
based on naive dimension counting arguments, the dynamics does indeed
drive an asymmetric evolution yielding three large dimensions. In
\cite{Easther:2003dd} we went further and studied the coupled
Einstein-Boltzmann equations for a thermal brane gas and found that
despite the naive dimension counting arguments, only highly
specialized initial conditions yield the desired brane wrapping
configuration.  In particular, we found that the spatial expansion
driven by the brane gas is generically too fast for brane interactions
to generate the expected anisotropies; instead, the branes quickly
freeze out.  However, this analysis still held out the possibility of
a loop-hole that would allow one to evade this discouraging
conclusion.  In the string theory corner of M-theory moduli space (the
very scenario studied in Brandenberger and Vafa's initial paper
\cite{BrandenbergerVafa}), the growth of spatial dimensions appeared
to be slower, perhaps allowing sufficient string-string interactions
to yield the desired asymmetric winding configuration and hence
asymmetric expansion. The main purpose of this paper is to study this
possibility in detail.

In section 2 we set up the basic framework of dilaton gravity, and in
section 3 we discuss the equilibrium thermodynamics of a string gas.
In section 4 we introduce the Boltzmann equations which govern string
annihilation and give a preliminary discussion of the phenomenon of
freeze-out. In section 5 we describe our method for sampling from the
possible initial states of the universe, and discuss holographic
bounds on the space of initial conditions.  In section 6 we present
numerical results, and show that other than for a narrow range of
initial conditions we get ``all'' or ``nothing'' evolution: either
there are too few strings to keep any spatial dimensions small, or
string interactions freeze out so quickly that a large number of
wrapped strings survive and prevent any dimensions from growing large.

\section{Dilaton gravity}

We start with type II string theory compactified on a 9-torus, with metric
\be
ds^2 = - dt^2 + \alpha' \sum_{i=1}^9 e^{2 \lambda_i(t)} d\theta_i^2
\qquad 0 \leq \theta_i \leq 2\pi\,.
\ee
The action for dilaton gravity is
\be
\label{dilgravact}
S = {1 \over 2 \kappa_{10}^2} \int d^{10} x \sqrt{-g} \, e^{-2\phi}
\left(R + 4 (\partial \phi)^2 + \cdots \right)
\ee
where (Polchinski \cite{Polchinski}, 13.3.24) $\kappa_{10}^2 = {1
\over 2} (2 \pi)^7 (\alpha')^4$.  From now on we set $\alpha' = 1$.
Following Tseytlin \& Vafa \cite{TseytlinVafa}, we define the shifted dilaton
\be
\label{ShiftedDef}
\varphi = 2 \phi - \sum_i \lambda_i
\ee
so that the action reads
\be
S = (2 \pi)^2 \int dt \, e^{-\varphi}\Bigl(\sum_i \dot{\lambda}_i^2
- \dtphi{}^2\Bigr)\,.
\ee
When one couples dilaton gravity to a matter system the time-time
component of the Einstein equations yields the Hamiltonian
constraint (or Friedman equation)
\be
\label{Hamiltonian}
(2\pi)^2 e^{-\varphi}\Bigl(\dtphi{}^2 - \sum_i \dot{\lambda}_i^2\Bigr) = E
\ee
where $E$ is the total matter energy.  This constraint implies that
$\dtphi{}^2$ never vanishes; we choose the direction of time so that
$\dtphi < 0$.  The dilaton equation of motion is
\be
\label{dilatoneom}
\ddot{\varphi} = {1 \over 2} \Bigl(\dtphi{}^2 + \sum_i \dot{\lambda}_i^2\Bigr)\,.
\ee
The scale factors obey
\be
\label{lambdaeom}
\ddot{\lambda}_i - \dtphi \dot{\lambda}_i = {1 \over 8\pi^2} e^{\varphi} P_i
\ee
where the ``total pressure'' $P_i = - {\partial F \over \partial
\lambda_i}$ is obtained by varying the matter free energy with respect
to $\lambda_i$.  $P_i$ is equal to the ordinary pressure in the
$i^{th}$ direction times the spatial volume.

Note that T-duality in the $i$th direction takes the simple form
\[
\lambda_i \rightarrow - \lambda_i, \hspace{0.5in} \varphi \ {\rm invariant}\,.
\]
This leaves the dynamics unchanged, provided that $E$ is invariant and
$P_i$ changes sign.

To get oriented we first study the vacuum equations, with all
pressures set to zero.  The equations of motion reduce to
\bea
\label{phieom}
&&\ddot{\varphi} = {1 \over 2} \Bigl(\dot{\varphi}{}^2 + \sum_i
\dot{\lambda}_i^2\Bigr) \\
\label{lameom}
&&\ddot{\lambda}_i - \dot{\varphi} \dot{\lambda}_i = 0\,.
\eea
Besides the trivial solutions in which the dilaton and radii are
constant, a Kasner-like branch of solutions can be obtained as
follows.  If the pressures vanish the energy $E$ is conserved.  Then
(\ref{phieom}) can be reduced to an equation just for $\varphi$, with
general solution
\be
\varphi(t) = \log{\left[\frac{16 \pi^2/E}{t(t+C)}\right]}
\label{k1}
\ee
(we have suppressed one constant of integration corresponding to an
arbitrary shift in $t$).  One can then integrate the $\lambda_i$
equations of motion to find
\be
\lambda_i(t) = A_i + B_i \log {t \over t + C}\,.
\label{k2}
\ee
The constants of integration $A_i$ are arbitrary, while in order to
satisfy the Hamiltonian constraint $B_i$ and $C$ must satisfy $C^2
\left(1-\sum_i B_i^2 \right) = 0$.  Thus either $C=0$ and the radii
are static, or $\sum_i B_i^2 = 1$ and the radii are time dependent.
In both cases, the dilaton rolls monotonically towards weak coupling.
  
We now turn to the the late-time asymptotic behavior of solutions to
the dilaton-gravity equations.\footnote{A similar analysis was
performed for M-theory in \cite{Easther:2002qk}.} First suppose the
pressure is negligible, $P_i \approx 0$, as is the case for a universe
in equilibrium with all radii sufficiently close to the self-dual
radius.  At late times the universe will approach the Kasner-like
solution (\ref{k1}), (\ref{k2}), with the asymptotic behavior
\be
\label{pressureless}
e^{\varphi} \sim {{\rm const.} \over t^2} \qquad \qquad e^{\lambda_i} \sim {\rm
const.}
\ee
Thus if $P_i \approx 0$ the dilaton rolls monotonically while the
radii approach constants in string frame.\footnote{This was also shown
in section 5 of \cite{Easther:2002qk}.  To relate the two solutions
note that \cite{Easther:2002qk} worked in terms of M-theory time
$t_M$, related to the string-frame time used here by $t_S \sim
t_M^{3/4}$.}

Now suppose that at late times we have $m$ unwrapped dimensions $x^1
\cdots x^m$ and $9-m$ wrapped dimensions $x^{m+1} \cdots
x^9$.\footnote{A dimension $x^i$ is called unwrapped if $\lambda_i >
0$ and the number of winding strings vanishes, or if $\lambda_i < 0$
and the number of momentum modes vanishes.}  Without loss of
generality we can go to a T-dual frame where the unwrapped
dimensions are all larger than string scale.  In this frame we expect
that at late times the universe will be dominated by a radiation gas
in the unwrapped dimensions; the pressures should vanish in the wrapped
dimensions due to a cancellation between winding and KK modes.  That
is, we expect
\be
P_i \sim \left\lbrace
\begin{array}{ll}
e^{-\lambda_i} & \quad i = 1,\ldots,m \\
0 & \quad i = m+1,\ldots,9
\end{array}
\right.
\ee
An ansatz which captures the late-time behavior is
\be
e^{\varphi} \sim {1 \over t^\alpha} \qquad
e^{\lambda_i} \sim \left\lbrace
\begin{array}{ll}
t^\beta & \quad i = 1,\ldots,m \\
{\rm const.} & \quad i = m+1,\ldots,9
\end{array}
\right.
\ee
Plugging this ansatz into the equations of motion fixes
\be
\alpha = {2 m \over m+1} \qquad\quad \beta = {2 \over m + 1}\,.
\ee
The dilaton rolls monotonically to weak coupling, while the unwrapped
dimensions grow with time and the wrapped dimensions have fixed sizes.
Thus if the string winding dynamics in the early universe favors $m =
3$, as suggested by the dimension counting argument reviewed in the
introduction, one could naturally explain why three spatial dimensions
become large.

\section{Equilibrium thermodynamics}
\label{Thermo}

In a coupled matter/gravity system the matter energy is determined by
the Hamiltonian constraint (\ref{Hamiltonian}).  For stringy matter
Tseytlin \& Vafa \cite{TseytlinVafa} long ago presented a simple
picture of the corresponding thermodynamics which is suitable for our
purposes.  There are two possible phases.

\subsection{Hagedorn phase}

Strings have a limiting Hagedorn temperature \cite{Hagedorn:st}.  For
weakly-coupled type II strings the limiting temperature is $T_H = {1
\over \pi \sqrt{8}}$.  Near this temperature the canonical ensemble
fails and one must use the microcanonical ensemble
\cite{Atick:1988si}.  In the Hagedorn phase the universe contains a
dense gas of winding and KK modes.  To a good approximation the free
energy $F = E - T_H S$ vanishes, so the microcanonical entropy is
given by $S(E) = E / T_H$.  The total pressure also vanishes, $P_i = -
{\partial F \over \partial \lambda_i} = 0$.

The thermodynamics of strings in the Hagedorn phase has been studied
by Deo, Jain and Tan \cite{DeoJainTan, DeoJainTan2}.  They employ the
microcanonical ensemble, with a fixed energy $E$ and a fixed net
winding charge in the universe; in a compact space the latter
vanishes.  They show that the average number of type II strings
present with winding charge vector ${\bf w}$ and energy $\epsilon$ is
given by
\be
D(\epsilon, {\bf w}, E) = \frac{N}{\epsilon} u(\epsilon, E)^{d/2}
e^{-u(\epsilon, E) {\bf w}^T A^{-1} {\bf w}/4}
\ee
where
\[ N = \frac{\left( 2 \sqrt{\pi} \right)^{-d} }{\sqrt{\det A}} \]
\[ u(\epsilon, E) = \frac{E}{\epsilon(E-\epsilon)} \]
\[ A_{ij} = \frac{1}{4 \pi^2 R_i^2} \delta_{ij} \]
Here $R_i \equiv e^{\lambda_i}$.  As a consistency check, note that
the total amount of energy in strings indeed adds up to $E$:
\[\int _0 ^E d \epsilon  \int d^dw\ \epsilon D(\epsilon, {\bf w}, E) = E. \]

We will ignore diagonally-wound topologies (where a string is
simultaneously wound on several dimensions) and assume that we have 9
unidimensional string gases.  That is, we set $d=1$ and assume
that the total energy available to each dimension is $1/9$ of the
total energy in the universe.  Thus the distribution for a single
winding charge $w_i$ is given by
\be
D(\epsilon,w_i,E) =  \frac{\sqrt{ \pi R}}{\epsilon} \sqrt{u(\epsilon,E/9)} 
  \exp{\left[-u(\epsilon,E/9) w_i^2 \pi^2 R^2_i \right]}.
\ee
The (thermally averaged) total number of positive windings $W_i$ is
then given by
\be
\label{wdist}
\langle W_i \rangle  = \int_0^{E/9} d\epsilon \int_0^\infty dw_i \ w_i D(\epsilon,w_i,E)
= \frac{\sqrt{E}}{12 \sqrt{\pi} R_i}.
\ee
Note that in (\ref{wdist}) we are only counting positive winding in
some direction; the net winding is zero.  The total length of string
present is proportional to the energy $E$, while the physical
distribution of string on the torus amounts to a random walk.  Thus
the average positive winding is simply the average distance from the
origin of a random walk.  This goes as the square root of the length,
or equivalently the square root of the energy.

We have computed the average number of winding modes, but a similar
result holds for the average amount of positive KK momentum present,
just by replacing $R_i \rightarrow 1/R_i$:
\be
\label{ndist}
\langle N_i \rangle = \frac{\sqrt{E}R_i}{12 \sqrt{\pi}}.
\ee

\subsection{Radiation phase}

Below the Hagedorn temperature the string oscillators make a
negligible contribution to the partition function, so we can focus on
single-string states which are labeled by an integer-valued momentum
vector $n_i$ and an integer-valued winding vector $w_i$.  In the
absence of a $B$-field the string energy levels are
\[
\epsilon({\bf n},{\bf w}) = \sqrt{\sum_i{\left(\left(\frac{n_i}{R_i}\right)^2 +
\left(w_i R_i\right)^2   \right)}}\,.
\]
The corresponding free energy for a gas of strings is
\[
\beta F = 128 \sum_{{\bf n} \cdot {\bf w} = 0} \log \tanh (\beta \epsilon({\bf n},{\bf
w})/2)
\]
where we have taken into account that for the type II string we have
128 bosonic and 128 fermionic species of excitations.  The condition
${\bf n} \cdot {\bf w} = 0$ enforces level matching.  For dimensions
which are large compared to a thermal wavelength we can approximate
momentum sums by integrals and neglect winding; likewise for
dimensions which are small compared to an inverse thermal wavelength
we can approximate winding sums by integrals and neglect momentum.
Any remaining intermediate-sized dimensions are frozen, with no
excitations.  Thus we have an approximate expression for the free
energy
\be
\label{ZeroModeBetaF}
\beta F \approx 128 \prod_{\rm large} 2 \pi R_i \prod_{\rm small} {2 \pi \alpha' \over R_i}
\int {d^d p \over (2 \pi)^d} \log \tanh (\beta \vert p \vert / 2)
\ee
where $d$ is the total number of unfrozen dimensions.  At this point
it is convenient to order
\be
\vert \lambda_1 \vert > \vert \lambda_2 \vert > \cdots > \vert \lambda_9 \vert
\ee
and to define the T-duality invariant `volume' of the $d$ unfrozen
dimensions
\be
\label{tvol}
V_d = \prod_{i=1}^d 2 \pi e^{\vert \lambda_i \vert} \,.
\ee
Then (\ref{ZeroModeBetaF}) can be identified with the free energy of a
massless ideal gas in $d$ spatial dimensions in a box of volume
$V_d$.
To write an equation of state we use the fact that in $d$ spatial
dimensions an ideal gas has an energy density $\rho = c_d T^{d+1}$,
where (for 128 bosonic and 128 fermionic degrees of freedom)
\be
c_d = 128 \cdot {2 d! \zeta(d+1) \over (4 \pi)^{d/2} \Gamma(d/2)} (2 - 2^{-d})\,.
\ee
The energy, entropy and total pressure of the gas are given by
\beas
E & = & c_d V_d T^{d+1} \\
S & = & {d + 1 \over d} c_d V_d T^d \\
P_i & = & - {\partial F \over \partial \lambda_i} = \left\lbrace
\begin{array}{ll}
{\rm sign}(\lambda_i) E / d & \quad i = 1,\ldots,d \\
0 & \quad i = d+1,\ldots,9
\end{array}
\right.
\eeas
However we still need to determine $d$.  If the energy is very small
then all dimensions are frozen.  As the energy increases
$\lambda_1,\lambda_2,\ldots$ will successively unfreeze.  This means
that the temperature in the radiation phase is given by
\be
\label{radtemp}
T_{\rm rad} = \min_k \left({E \over c_k V_k}\right)^{1/(k+1)}\,.
\ee
The value of $k$ which minimizes the right hand side is equal to the
number of unfrozen dimensions.  $T_{\rm rad}$ calculated in this way
could be larger than the Hagedorn temperature; this signals that the
system is actually in the Hagedorn phase.  That is the true
temperature of the system is $\min (T_{\rm rad}, T_H)$.

In the radiation phase we can compute the amount of positive KK
momentum present in equilibrium by using the fact that for a
one-dimensional massless gas $\langle N_i \rangle$ is related to the
pressure by
\be
\label{radn}
\langle N_i \rangle = \frac{1}{2} P_i R_i .
\ee
This estimate makes sense for $R_i \gg \sqrt{\alpha'}$, in which case
we also have $\langle W_i \rangle = 0$.  If on the other hand $R_i \ll
\sqrt{\alpha'}$ we just use the T-dual formulas
\begin{eqnarray}
\label{radw}
\langle W_i \rangle &=&- \frac{1}{2} P_i / R_i \\
\nonumber
\hspace{0.5in} \langle N_i \rangle &=& 0.
\end{eqnarray}

\section{Winding and KK Annihilations}
\subsection{Boltzmann equations}
If the universe was in equilibrium we could just insert the results of
the previous section into the dilaton-gravity equations of motion.
But the Brandenberger-Vafa scenario is driven by departures from
thermal equilibrium.  A crude way to keep track of these departures is
to let $W_i$ be the amount of positive winding charge around dimension
$i$.  Likewise let $N_i$ be the amount of positive Kaluza-Klein
momentum in direction $i$.  Of course since the space is compact there
must also be $W_i$ units of anti-winding charge and $N_i$ units of
anti-KK-momentum.

Let us, for the time being, assume that each unit of charge is carried
by a single string: that is, there are $W_i$ strings each wound once
with positive orientation, similarly for the KK modes.  We will also,
for the time being, assume that the strings have no oscillator
excitations.  Then the annihilation of these momentum and winding
modes is governed by Boltzmann equations, similar to the equations
that govern the evolution of M2-brane winding \cite{Easther:2003dd}:
\begin{eqnarray}
\label{singleboltz}
{d N_i \over dt} & = & - {\langle f(v) \rangle \over 2 \pi} e^{\varphi - 2 \lambda_i}
(N_i^2 - \langle N_i \rangle^2) \\
\nonumber
{d W_i \over dt} & = & - {\langle f(v) \rangle \over 2 \pi} e^{\varphi + 2 \lambda_i}
(W_i^2 - \langle W_i \rangle^2)
\end{eqnarray}
The cross-section for bosonic wound strings was calculated by
Polchinski \cite{Polchinski:cn}, who found $f(v) = 2/(1-v^2)$ for two
anti-parallel strings moving with velocity $v$.  Subsequent studies
have evaluated this quantity for F- and D-strings
\cite{Jackson:2004zg}.  We will set $f(v) \approx 2$, appropriate for
a gas of slowly-moving strings.

Let us make a few comments on the structure of these Boltzmann
equations.  First, note that they are invariant under T-duality.
Second, note that they respect the dimension-counting arguments of
Brandenberger \& Vafa \cite{BrandenbergerVafa}.  An implicit factor of
the inverse volume of the universe is present in the definition of
$e^\varphi$ (\ref{ShiftedDef}).  But due to the factor $e^{2
\lambda_i}$ upstairs in the equation for $d W_i / dt$, strings wrapped
on a large 3-torus will still be able to annihilate effectively, just like
particles moving in one large spatial dimension.  Finally, we should
contrast our Boltzmann equations with the results presented in
\cite{Brandenberger:2001kj}, which were appropriate for strings with a
dilaton-independent cross-section such as cosmic strings.

Continuing our study of the Boltzmann equations, we now include string
oscillator excitations but still restrict attention to unit winding
and momentum charges. With oscillators excited a more accurate
cross-section is obtained by replacing\footnote{We thank R.\ Myers for
bringing this issue to our attention.}
\[ \exp(-\lambda_i)  \rightarrow  \epsilon_i \]
in the Boltzmann equation for $dN_i/dt$, where $\epsilon_i$ is the
average energy of a string with a unit of KK momentum.  Likewise in
the $dW_i/dt$ equation we should replace
\[ \exp(+\lambda_i)  \rightarrow  \delta_i \]
where $\delta_i$ is the average energy of a unit winding string.
This is supported by the results of Lizzi and Senda
\cite{Lizzi:1990nc}, who redo Polchinski's calculation for two highly
excited strings -- which have many oscillator excitations but no
winding -- and show that the interaction rate goes like the product of
the energies of the two strings.  This modifies the Boltzmann
equations to read
\begin{eqnarray}
\nonumber
{d N_i \over dt} & = & - {1 \over \pi} e^{\varphi } \langle \epsilon_i \rangle^2
(N_i^2 - \langle N_i \rangle^2) \\
\label{energyboltz}
{d W_i \over dt} & = & - {1 \over \pi} e^{\varphi } \langle \delta_i \rangle^2
(W_i^2 - \langle W_i \rangle^2).
\end{eqnarray}
We may estimate the typical energy per momentum or winding mode from
the distributions given in the previous section:
\begin{eqnarray}
\nonumber
{\rm Hagedorn \ phase} &:& \langle \epsilon_i \rangle = \frac{E}{9\langle N_i \rangle} \quad \hbox{\rm (momentum modes)} \\
\label{StrandEnergy}
                       & & \langle \delta_i \rangle = \frac{E}{9\langle W_i \rangle} \quad \hbox{\rm (winding modes)}
\end{eqnarray}
\begin{eqnarray}
\nonumber
{\rm Radiation \ phase} &:& \langle \epsilon_i \rangle = 1/R_i \quad \hbox{\rm (momentum modes)} \\
\nonumber
                        & & \langle \delta_i \rangle = R_i \quad \hbox{\rm (winding modes)}\,.
\end{eqnarray}
Here we have assumed that in the Hagedorn phase the energy is equally
distributed between dimensions.  Note that in the Hagedorn phase the
average energy per mode scales as $\sqrt{E}$.

Finally we consider  strings that are multiply wound
around each dimension, so that the number of positively-wound strings
and the winding charge $W_i$ are not necessarily the same. We may
think of the winding charge as made up of $W_i$ open unit strands that
are braided together to form closed strings.  Depending on the
braiding, there can be anywhere from 1 to $W_i$ closed strings present.  Also
depending on the braiding, an individual closed string can carry anywhere
from 1 to $W_i$ units of winding charge.  There are $W_i!$ ways of
braiding the strands; we assume all braidings are equally likely.
Then the typical strand is part of a closed string that carries winding
charge $(W_i+1)/2$.  The cross-section of a string is proportional to
its length and hence winding charge, thus we expect the typical
string-string cross-section to be enhanced by a factor
$((W_i+1)/2)^2$.  Making this modification to the cross-section, and
rewriting the Boltzmann equation as an equation for the rate of change
of the positive winding charge, a net enhancement factor of
$(W_i+1)/2$ appears on the right hand side relative to
(\ref{energyboltz}).

With an analogous modification for multiple-momentum strings the
Boltzmann equations read
\bea
\label{multboltz}
{d N_i \over dt} & = & - {(N_i+1) \over 2 \pi} e^{\varphi } \langle \epsilon_i \rangle^2
(N_i^2 - \langle N_i \rangle^2) \\
\nonumber
{d W_i \over dt} & = & - { (W_i+1) \over 2 \pi} e^{\varphi} \langle \delta_i \rangle^2
(W_i^2 - \langle W_i \rangle^2)
\eea
where the average energies per strand are still given by (\ref{StrandEnergy}).

One could question our assumption that all braidings are equally
likely.  Although this seems like a reasonable assumption when the
universe is small, on entropic grounds it could be that as some
dimensions grow large singly wound strings become favored.  Rather
than study this issue directly, in our numerical work we will
investigate the two extreme possibilities: all strings singly-wound as
in (\ref{energyboltz}), or all braidings equally likely as in
(\ref{multboltz}).

\subsection{Freeze-out}
In an expanding universe the evolution of a species depends on the
species' annihilation rate $\Gamma$ and the cosmological expansion
rate (or Hubble parameter) ${\dot R}/R = {\dot \lambda}$.  For dilaton
gravity one also needs to take into account the rate of change of the
dilaton $\dot{\varphi}$.  To determine whether a nonzero number of
strings survive to the asymptotic future, we need to study how these
parameters evolve. If the annihilation rate of wound strings decreases
too rapidly it could undermine the naive dimension counting
arguments, which implicitly assume that $\Gamma$ remains non-zero.

To illustrate the possibility of freeze-out consider the following
simple situation. At some initial time set all $\lambda_i = 0$.  Introduce
the same number of unit winding and unit momentum strings in
all directions: $N_i = W_i \equiv N$.  Suppose further that no
oscillators are excited.  Then the pressures $P_i$ all vanish, and it
is consistent to set the logarithmic scale factors $\lambda_i = 0$ for
all time. The remaining equations of motion are very simple.  As
explained in section 2, the shifted dilaton obeys $\ddot{\varphi} = {1
\over 2} \dot{\varphi}^2$ with solution
\[
e^{\varphi} = {A \over (t - t_0)^2}.
\]
Here $A$ and $t_0$ are two constants of integration.  As expected the
dilaton rolls monotonically to weak coupling.  The Boltzmann equation
(for singly-wound strings) reads
\[
{dN \over dt} = - {1 \over \pi} e^{\varphi} \langle \epsilon \rangle^2
\left(N^2 - \langle N \rangle^2\right)\,.
\]
To get a feel for whether the strings will freeze out it suffices to
set $\langle N \rangle = 0$.  Then the general solution is
\begin{equation}
\label{BoltzmannSolution}
{1 \over N(t)} = {1 \over N(t_1)} + {1 \over \pi} \int_{t_1}^t dt' \,
e^{\varphi} \langle \epsilon \rangle^2\,.
\end{equation}
As long as the integral stays finite as $t \rightarrow \infty$ a
non-zero fraction of the strings will freeze out.  In the case at hand
the pressure vanishes, which means the total energy in matter does not
change with time; since the radii are fixed the average energy per
string $\langle\epsilon\rangle$ also remains constant.  Then the
integral is strongly convergent, and
\[
{1 \over N(t)} = {1 \over N_\infty} - {A \langle \epsilon \rangle^2 \over \pi (t -
t_0)}\,.
\]
Here we have defined $1/N_\infty \equiv 1/N(t_1) + A \langle \epsilon
\rangle^2 / \pi (t_1 - t_0)$.  As $t \rightarrow \infty$ a non-zero
fraction of the unit winding and unit momentum strings do freeze out,
with $N(t) \rightarrow N_\infty$.

This sort of behavior should be fairly generic, even for solutions
that do not sit precisely at the self-dual radius. The pressure
vanishes as long as one remains in the Hagedorn phase, giving give
rise to a conserved matter energy.  Moreover, if the radii change
slowly with time, the average energy per string still remains roughly
constant.  The dilaton, however, will still roll monotonically towards
weak coupling, and as long as it does so quickly enough for the
integral in (\ref{BoltzmannSolution}) to converge, some strings will
freeze out.  The enhanced cross-section due to multiple winding in
(\ref{multboltz}) does not change this outcome.

This is troubling for the Brandenberger-Vafa scenario, as it shows
that simple dimension-counting arguments can fail to capture the true
dynamics of winding strings.  Our goal in the remainder of this paper
is to undertake a detailed numerical investigation of the likelihood
of freeze-out.

\section{Initial conditions and Holography}

We would like to choose initial conditions at random, so as to
uniformly sample the possible states of the early universe. In
practice we proceed by fixing the initial value of the shifted dilaton
$\varphi$ and the initial volume of the universe $V$.  All other
degrees of freedom will be given random initial values, drawn from the
probability distribution worked out below.

Ideally, we would  average over all possible values of the
``coordinates" $\lambda_i, \varphi$ together with their canonical
momenta using the Liouville measure obtained from the action
(\ref{dilgravact}).  The microcanonical volume of phase space for
dilaton gravity plus matter is
\be
\label{Omega}
\Omega \sim \int d^9\lambda \, d^9\dot{\lambda} \, d\varphi \, d\dtphi \,\,
e^{-10 \varphi} e^S
\ee
where $S$ is the matter entropy.  In the Hagedorn phase this is given by
\be
\label{ProbDist}
S = E / T_H = (2 \pi)^2 e^{-\varphi} (\dtphi{}^2 - \sum_i \dot{\lambda}_i^2) / T_H \,.
\ee
Thus at the level of supergravity the initial conditions which maximize the
entropy are
\be
\begin{array}{ll}
\varphi \rightarrow - \infty & \qquad
\hbox{\rm weak string coupling and large volume} \\*[4pt]
\dtphi \rightarrow - \infty & \qquad
\hbox{\rm effective coupling rapidly decreasing} \\*[4pt]
\dot{\lambda}_i = 0 & \qquad
\hbox{\rm constant size of torus}
\end{array}
\ee
To set initial conditions we first fix a value of $\varphi$.  In order
for effective supergravity to be valid we must have $e^\varphi \ll 1$.
Note that since we're working with effective supergravity, not string
theory, only the value of the shifted dilaton matters and we don't
need to worry about the underlying dilaton $\phi$ defined in
(\ref{ShiftedDef}) becoming large.  Next we fix a value for ${\dot
\varphi}$.  For supergravity to be valid we must have ${\dot \varphi}
\gtrsim -1$.  Then from (\ref{ProbDist}) note that the ${\dot
\lambda}$ are Gaussian distributed, with a characteristic spread
\be
(\Delta {\dot \lambda}_i)^2 = T_H (2 \pi)^{-2} e^{\varphi} \,.
\ee
For simplicity we take the ${\dot \lambda}_i$'s to be uniformly
distributed about zero, in the interval $[-\sqrt{T_H} (2 \pi)^{-1}
e^{\varphi/2}, \sqrt{T_H} (2 \pi)^{-1} e^{\varphi/2}]$.  Note that for
$\varphi < 0$ we'll have $-1 < {\dot \lambda}_i < 1$.  The scale
factors $\lambda_i$ do not appear in the entropy, so we take them to
be uniformly distributed, subject only to the constraint that the
T-duality invariant 9-volume $V$ defined in (\ref{tvol}) has the
specified value: $\sum_i \vert \lambda_i \vert = \log (V/(2\pi)^9)$.

Following the M-theory analysis of \cite{Easther:2003dd}, we can ask
about the holographic bound
\cite{'tHooft:gx,Susskind:1994vu,Fischler:1998st,Easther:1999gk}.
This is
\be
S \leq \frac{A_E}{4G} = \frac{2 \pi A_E}{\kappa^2} =
\frac{2 \pi \Omega_8 R_E ^8}{\frac{1}{2} (2 \pi)^7 (\alpha')^4}
\ee
where $\Omega_8$ is the area of a unit $S^8$ and the subscript $E$
reminds us that this must be calculated in the Einstein frame.  We
wish to convert this to string frame, with (note that the regular
dilaton, not $\varphi$, appears below)
\be
R_E = e^{-\phi/4} R_S
\ee
making
\be
S \leq \frac{2 \Omega_8 \pi^8}{(2 \pi)^6} e^{-\varphi - \lambda}
\ee
on an isotropic torus.  Comparing to the initial value we get a bound
(again assuming we start in the Hagedorn phase)
\be
S = \pi \sqrt{8} (2 \pi)^2 e^{-\varphi} \left( {\dot \varphi}^2 - \sum_i {\dot
\lambda}_i \right) \leq \frac{2 \Omega_8 \pi^8}{(2 \pi)^6} e^{-\varphi
- \lambda}
\ee \be
\pi \sqrt{8} (2 \pi)^2 {\dot \varphi}^2 \leq \frac{2 \Omega_8 \pi^8}{(2 \pi)^6}
e^{-\lambda}
\ee \be
2 \pi e^\lambda \leq \frac{\Omega_8}{128 \sqrt{2} {\dot \varphi}^2 }
\ee
We can interpret this as a bound
\be
V \sim (2 \pi e^\lambda)^9 \leq \left( \frac{\Omega_8}{128 \sqrt{2}
{\dot \varphi}^2 } \right)^9 = \left( \frac{\pi^4}{420 \sqrt{2} {\dot \varphi}^2 }
\right)^9
\ee
or equivalently
\be
\label{HoloBound}
\dot{\varphi}^2 \leq { 0.16 \over V^{1/9}}\,.
\ee

\section{Numerical Analysis}

\subsection{Initial Conditions} 

Our simulations proceed by generating multiple sets of initial data,
solving the equations of motion numerically, and looking at the number
of wrapped dimensions at late times after freeze-out has taken place.
In each run we fixed the initial values of $V$ and $\varphi$.  For the
most part we started with $\dot{\varphi} = -1$; this maximizes the
entropy while keeping supergravity valid.\footnote{Note that we are
considering the case where the dilaton is rolling towards weak
coupling.}  All other initial conditions are allowed to fluctuate
randomly.  The $\dot{\lambda}_i$ are chosen from the flat distribution
described in the previous section.  The initial $\lambda_i$ are
generated by choosing nine random numbers in $[-1,1]$ and applying an
overall scaling so that the initial volume matches the specified
value. To assign initial values to $N_i$ and $W_i$ we compute the mean
values from section 3 and then add random thermal fluctuations about
the mean, of magnitude
\be
\label{noise}
\Delta N_i \approx \sqrt{\langle N_i \rangle} \qquad \Delta W_i \approx \sqrt{\langle W_i \rangle}\,.
\ee
Note that we do not impose the holographic bound (\ref{HoloBound}) on
our initial data; we have some comments on this below.

To evolve the system we use a Runge-Kutta algorithm.  At each time
step we begin by computing the total matter energy from the
Hamiltonian constraint (\ref{Hamiltonian}).  The equilibrium
thermodynamics discussed in section 3 enables us to decide whether the
system is in a Hagedorn or radiation phase.  Based on this we compute
the corresponding thermal expectation values $\langle N_i \rangle$,
$\langle W_i \rangle$.  We then use the equations of motion to evolve
to the next time step.

The dilaton-Einstein equations of motion are given in
(\ref{dilatoneom}), (\ref{lambdaeom}).  To solve them we need an
expression for the pressures $P_i$.  We set\footnote{Here we are
relating the actual pressures $P_i$ to the actual values of $N_i$ and
$W_i$, allowing for departures from thermal equilibrium.  Thus
(\ref{OffEquil}) should not be confused with (\ref{radn}),
(\ref{radw}) where we used the equilibrium pressures to compute the
thermally averaged values of $N_i$ and $W_i$.  Of course for a system
in equilibrium the expressions are compatible.}
\be
\label{OffEquil}
P_i = 2 \left( N_i e^{-\lambda_i} - W_i e^{\lambda_i} \right)\,.
\ee
This simple estimate is valid when no oscillators are excited; thus it
should be accurate in the radiation phase.  In the Hagedorn phase
oscillators are excited and the pressure receives corrections.
However given the equilibrium values of $N_i$ and $W_i$ in the
Hagedorn phase (\ref{wdist}), (\ref{ndist}) note that on average a
cancellation makes the pressure vanish.  Thus, although it would be
nice to have a more precise expression for the pressure, we do not
expect any refinements to (\ref{OffEquil}) to significantly affect our
results.

The Boltzmann equations were discussed in section 4.  To allow for the
effects of multiply-wound strings we ran simulations using two
different versions of the Boltzmann equations given in
(\ref{energyboltz}) and (\ref{multboltz}).  The first is appropriate
for strings that only carry one unit of winding or momentum charge,
while the second is appropriate for strings with multiple winding or
momentum charges.

In Figure 1 we show what happens when we vary the initial values of
$\varphi$ and $V$, starting with the initial condition $\dot{\varphi}
= -1$.  We show the average number of wrapped dimensions present both
in the initial configuration and after freeze-out.  Clearly a final
state with three unwrapped dimensions is not dynamically favored.  If
one begins at reasonably strong coupling then very few strings are
present in the initial state, while if one begins at weak coupling
string interactions turn off too rapidly for the required
annihilations to occur.  In either case three large spatial dimensions
is not the most likely late-time geometry.

We have explored what happens if the initial value of $\dot{\varphi}$
is decreased, since the holographic bound (\ref{HoloBound}) restricts
the allowed values of this quantity.  In Figure 2 we show the initial
and final number of wrapped dimensions starting with the initial
condition $\dot{\varphi} = -0.15$.  The qualitative outcome is the
same, just shifted to more negative initial values of $\varphi$.  This
is not surprising, given the Hamiltonian constraint
(\ref{Hamiltonian}): roughly speaking a change in $\dot{\varphi}^2$
can be compensated by shifting $\varphi$ so as to keep the total
energy fixed.

In Figure 3 we show the distribution of initial winding configurations
for universes that end up with three large dimensions.  Although the
number of wrapped dimensions can either increase or decrease with
time, it is unlikely that one can begin deep in the Hagedorn phase
with nine wrapped dimensions and end up with a three dimensional
universe.

In Figure 4 we show how the distribution of final winding
configurations depends on the initial value of $\varphi$.  Although
for a rather narrow range of $\varphi$ three dimensions is the favored
outcome, the distribution of final dimensionality is not very sharply
peaked.

All results presented so far have been based on the multiply-wound
cross section (\ref{multboltz}).  We have studied what happens if we
evolve the system using the singly-wound cross section
(\ref{energyboltz}).  The change in the results is negligible, much
less than the widths of the distributions shown in Fig.~4.  Thus the
qualitative outcome is the same, with no dynamical preference for
three dimensions.

\section{Conclusions}
Our results indicate that -- within the context of our approximations
-- the expansion of the universe has an ``all or nothing"
character. If initial conditions are such that one begins with many
wrapped strings, the strings typically freeze out and keep all
dimensions small. On the other hand if one begins with few wrapped
strings, the strings typically annihilate and all dimensions
decompactify. Between these extremes there are initial conditions that
lead to three large dimensions, but such initial conditions are not
generic.  Fine-tuning the initial conditions to yield three large
dimensions is thus possible, but runs counter to the goal of the
string gas program: finding a mechanism in which generic initial data
yields three large spatial dimensions.

The unexpected chink in the Brandenberger-Vafa scenario we have found
is that due to the rolling dilaton the string annihilation cross
section becomes weaker than previously realised.  To avoid this
impasse we would need a mechanism for keeping the string annihilation
cross section sufficiently robust.\footnote{To study the long-time
behavior with an enhanced cross-section one should take thermal
fluctuations into account, not only in setting the initial conditions
as in (\ref{noise}), but also by adding an explicit noise term to the
equations of motion.} There are ways in which this might be
accomplished (e.g. strings wound on finite fundamental groups
\cite{Easther:2002mi}, in confining backgrounds \cite{Jackson:2004zg},
or having unusual kinematic configurations \cite{Mende:1994wf}), but
as yet none have been studied in adequate detail to determine their
viability.  Also we should note that, even if one manages to stabilize
the cross-section, one would still have to face the issue that the
gravitational back-reaction of an anisotropic string gas turns off at
late times, due to the factor $e^{\varphi}$ which appears in the
equations of motion (\ref{lambdaeom}).  One might be tempted to
postulate a mechanism which stabilizes the dilaton, however this is
problematic for reasons discussed in \cite{Berndsen:2004tj}: with pure
Einstein gravity strings should freeze out, along the lines of our
M-theory analysis \cite{Easther:2003dd}.

Lest we appear too pessimistic, let us note some directions for future
study which might invalidate our conclusions.
\begin{itemize}
\item
In this paper we have only studied decreasing dilaton solutions,
whereas there is also a class of increasing dilaton solutions. If the
value of the dilaton grew sufficiently large, the appropriate
framework would be M-theory, and the results of our previous paper
\cite{Easther:2003dd} would apply. However, it is possible that there
is an intermediate time in which the dilaton is large enough for
string annihilations to be effective, yet small enough for
perturbative string theory to be relevant.
\item
In this paper we made a number of simplifying assumptions.  In
particular we assumed spatial homogeneity and only considered the
radial moduli of the torus.  A more complete analysis at the level of
effective supergravity would be desirable; steps in this direction
have been taken in \cite{Watson:2003uw,Watson:2004vs,Campos:2004yn}.
\end{itemize}

The failure of the string gas scenario to naturally lead to three
large spatial dimensions may be telling us one of three things.
First, perhaps the string gas (or brane gas) framework is supplanted
by other dynamics in the early universe, invalidating the approach we
have been following. Second, perhaps the measure (\ref{Omega}) does
not reflect the true distribution of possible initial conditions of
the universe.  Third, perhaps three spatial dimensions is not favored.
If one imagines that many ``universes'' are created, all with
different initial conditions, then some sort of anthropic argument
could be invoked.  But many people, including ourselves, are
uncomfortable with anthropic arguments until every other possibility
has been explored. Consequently we intend to return to these
cosmological issues as our understanding of string theory in the early
universe improves.

\section{Acknowledgements}
BG and DK are supported in part by DOE grant DE-FG02-92ER40699. RE is
supported in part by DOE grant DE-FG02-92ER-40704.  MGJ and DK are
supported in part by US--Israel Binational Science Foundation grant
\#2000359.  ISCAP gratefully acknowledges the financial support of the
Ohrstrom Foundation.  We would like to thank the organizers of the
``Superstring Cosmology'' conference at KITP and the ``New Horizons in
String Cosmology'' conference at Banff.  DK would like to thank
Rutgers University for their hospitality.  We wish to thank R.\
Brandenberger, D.\ Easson, S.\ Kachru, R.\ Myers and L.\ Susskind for
valuable discussions.


\begin{figure}
\centerline{\epsfig{file=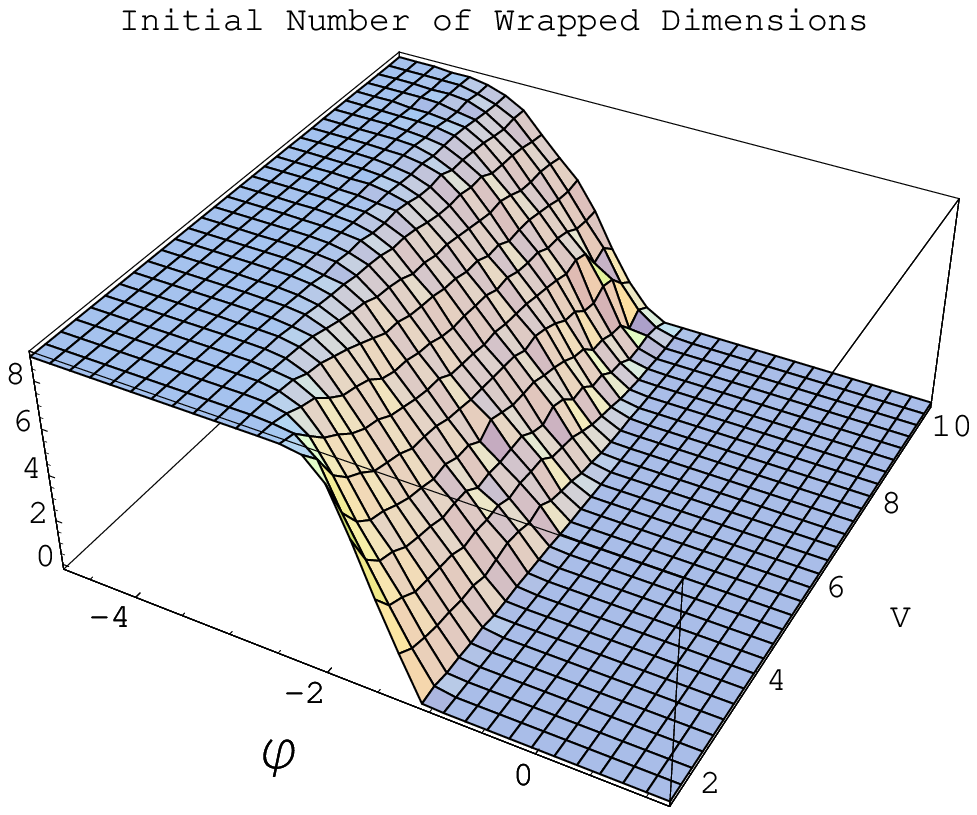,height=210pt}}
\vspace{1.0cm}
\centerline{\epsfig{file=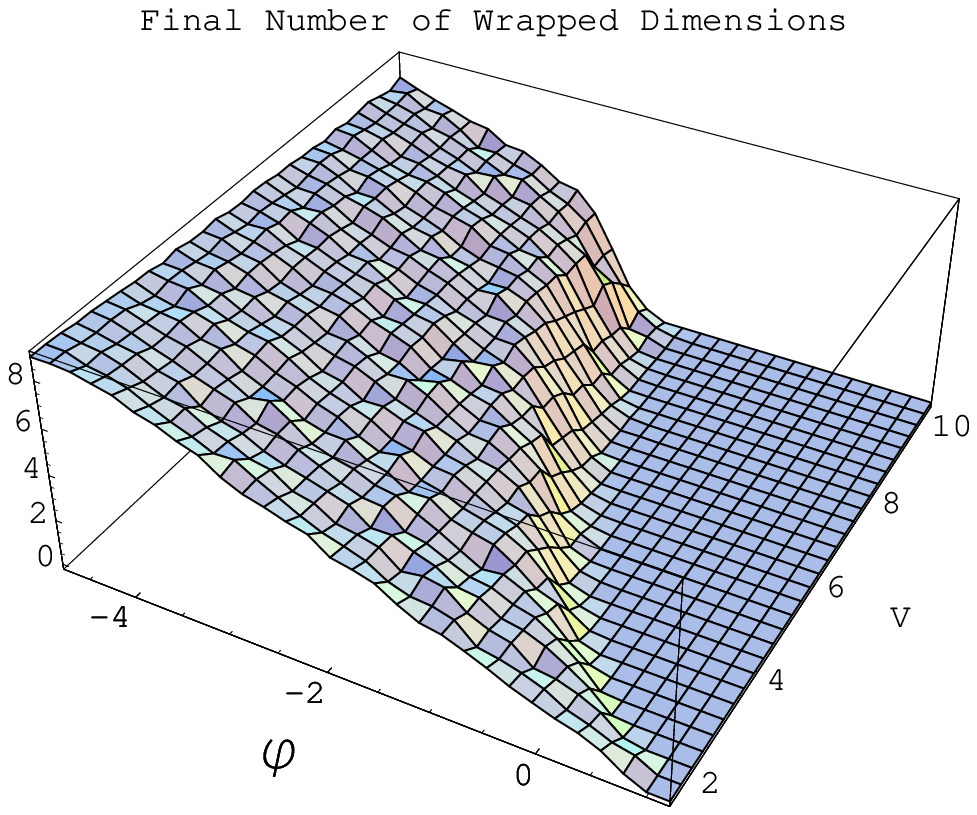,height=210pt}}
\caption{Average initial and final number of wrapped dimensions as a
function of the initial coupling and initial volume, starting with
$\dot{\varphi} = -1$ and evolved using the multiply-wound cross
section.  The volume is measured in units of $(2 \pi
\sqrt{\alpha'})^9$.}
\end{figure}

\begin{figure}
\centerline{\epsfig{file=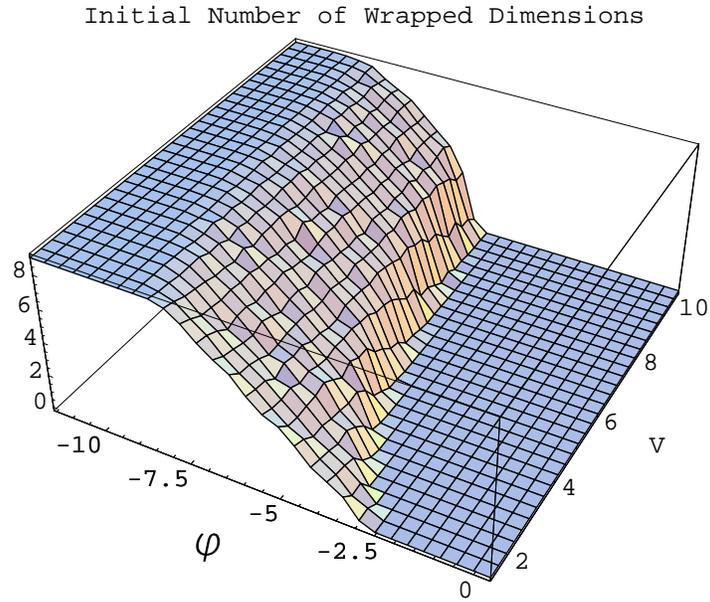,height=230pt}}
\vspace{1.0cm}
\centerline{\epsfig{file=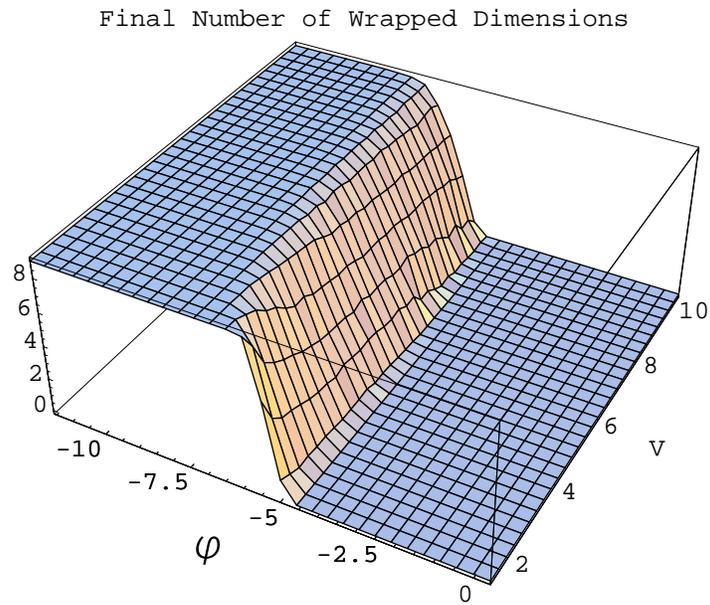,height=230pt}}
\caption{Illustrates the dependence on the initial value of
$\dot{\varphi}$.  Same as Fig.~1 except the simulations begin with
$\dot{\varphi} = -0.15$.}
\end{figure}

\begin{figure}
\centerline{\epsfig{file=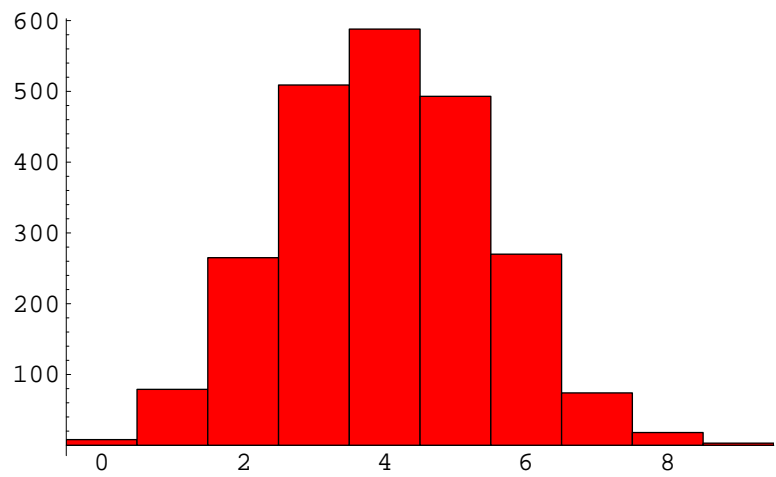}}
\caption{A histogram showing the distribution of the initial number of
unwrapped dimensions for universes that end up three dimensional.
Extracted from the data set used to generate Fig.~1.}
\end{figure}

\begin{figure}
\centerline{
\epsfig{file=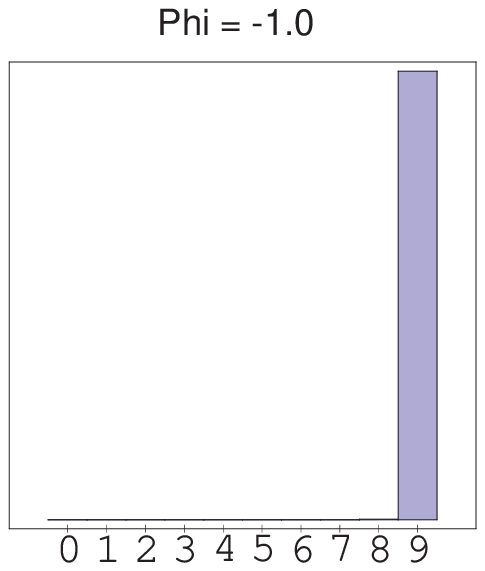}
\hspace{-2.7cm}
\epsfig{file=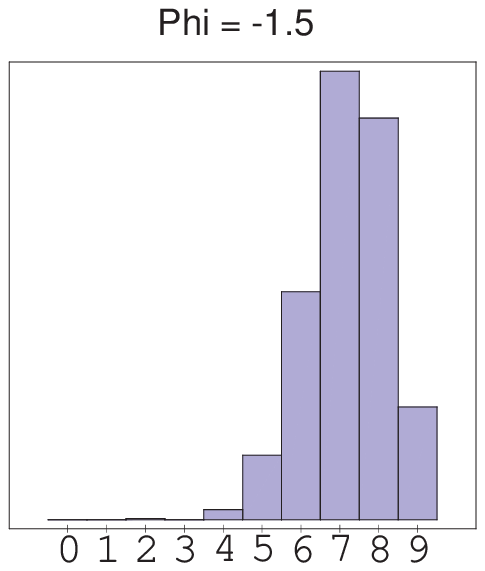}
\hspace{-2.7cm}
\epsfig{file=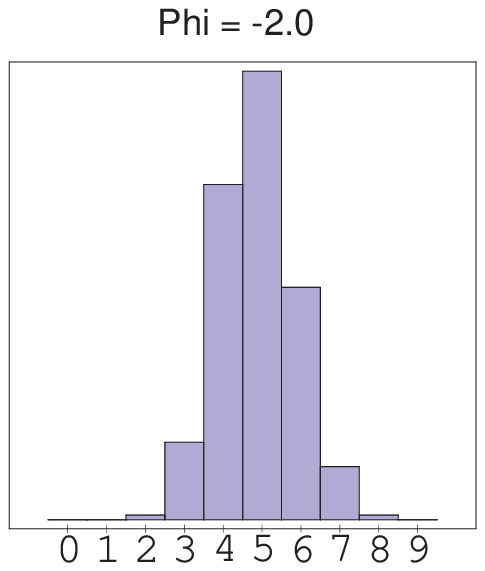}}
\vspace{-1.5cm}
\centerline{
\epsfig{file=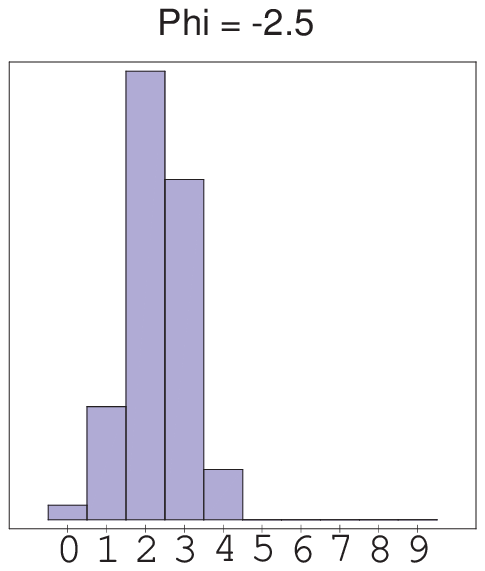}
\hspace{-2.7cm}
\epsfig{file=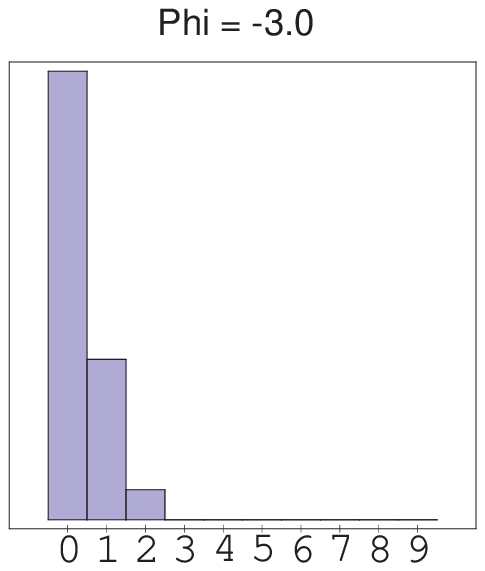}
\hspace{-2.7cm}
\epsfig{file=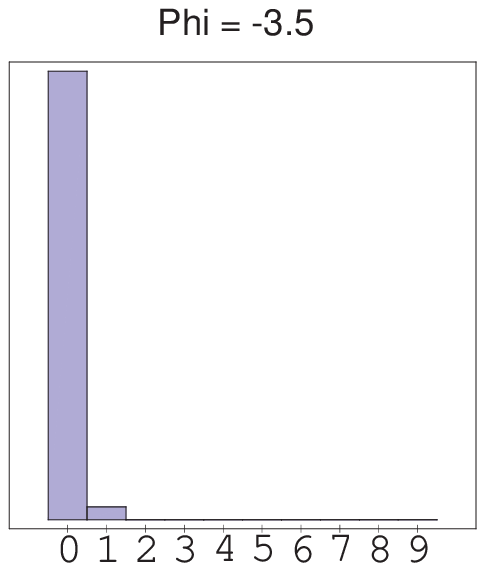}}
\caption{Histograms showing the distribution in the number of
unwrapped dimensions at late times for various initial values of
$\varphi$.  Each histogram is based on $10^3$ simulations at an
initial volume $V = 4.0 \times (2 \pi \sqrt{\alpha'})^9$ and an
initial $\dot{\varphi} = -1$.}
\end{figure}

\end{document}